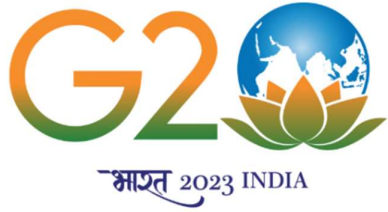 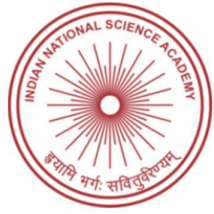 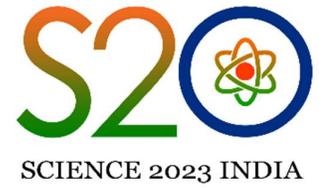

# Policy Brief

## LARGE PROJECTS IN ASTRONOMY: AN INDIAN ENDEAVOUR

Discussed and drafted during S20 Policy Webinar on Astroinformatics for Sustainable Development held on 6-7 July 2023

Contributors: Yogesh Wadadekar, Shriharsh Tendulkar, Sanjit Mitra, Eswar Reddy, A N Ramaprakash, Timo Prusti, Juna Kollmeier, Pranav Sharma, Ashish Mahabal

**Introduction**
Cutting-edge astronomy initiatives often entail substantial investment and require a high level of expertise. Even the most technologically advanced nations recognize the value of establishing international partnerships to secure both financial resources and talent for these ambitious endeavours. Key examples of such collaborations include groundbreaking projects like the Ground-based Next-Generation Optical and Infrared Telescopes (GMT, TMT, and E-ELT), the Square Kilometre Array (SKA) for radio astronomy, and the Laser Interferometer Gravitational-Wave Observatory (LIGO) for gravitational astronomy. These ventures hold immense promise for catalysing transformative scientific discoveries, driving technological innovation, provide training opportunities for the next generation of scientists and engineers, and expanding our understanding of the cosmos that surrounds us.

Crucially, large-scale multilateral collaborations serve as powerful agents for promoting unity and peace among the global population. Participants from various nations share a vested interest in the success of these projects and the wealth of knowledge they yield, fostering a sense of common purpose and shared goals.

By utilizing astroinformatics capabilities, these initiatives are not merely enhancing our comprehension of the universe but are also actively contributing to the attainment of sustainable development objectives. They are accomplishing this by fostering international cooperation, education, and technological progress. In this discussion, we delve into the challenges faced, accomplishments achieved, and prospective avenues for substantial astronomical undertakings. Additionally, we present recommendations aimed at guaranteeing their effectiveness and optimizing their influence on both scientific advancement and society.

**LIGO and LIGO-India**
LIGO-India is an Astronomy mega-science project on Indian soil. The project recently received approval from the Union Cabinet of India and aims to start observing by 2030. LIGO-India aims to operate at the same sensitivity as the two LIGO detectors in the US at the time of operation. It will significantly boost precise localisation of astrophysical events on the sky, which will, in turn, vastly improve the chances of following up those events using electromagnetic telescopes and the Astrophysics that we learn from those observations.

The observatory will be built by the Department of Atomic Energy (DAE) and the Department of Science and Technology (DST), Government of India, with a Memorandum of Understanding (MoU) with the National Science Foundation (NSF), USA, along with several national and international research and academic institutions. The project is being led by four institutions, Directorate of Construction Services and Estate Management (DCSEM), Institute of Plasma Research (IPR), Inter-University Centre for Astronomy and Astrophysics (IUCAA), and Raja Ramanna Centre for Advanced Technology (RRCAT). Being multidisciplinary in nature, a mega-science project like LIGO-India fosters collaborations across various expertise and creates a common platform for exchanging knowledge and expertise. Such platforms can immensely help in building foundations for sustainable development. The scientific requirement for LIGO-India demanded a site which is remote. This provides the opportunity to bring a major boost in education and awareness, especially the excitement for science, in the neighbouring areas, especially to the underprivileged population.

**SKA and SKA-India**
The Square Kilometre Array Observatory (SKAO) is an ambitious international project to build and operate the next generation radio telescope. To realize the goal of the SKAO, an intergovernmental organization along the lines of CERN and ESO has been set up with headquarters at Jodrell Bank, UK. To date, nine countries have ratified the convention and joined the SKAO. India has been participating in SKA developments actively since 2010 and our proposal for India to join SKAO is under consideration at the highest levels of the Indian government.

Indian participation in the SKA is coordinated by a consortium of universities and research institutions from all over the country called the SKA-India consortium. Scientists have formed their national working groups in various sub-fields of astrophysics. They also participate actively in the corresponding international working groups. Engineers at NCRA and RRI work actively with software industry partners and bespoke hardware vendors on various aspects of the telescope construction. In software, India is a major contributor to the observatory control and management systems. All work is done collaboratively with the SKAO as well as with industry and academic partners in all SKA member countries, which are almost all members of the G20. A distinguishing feature of the Indian participation in the SKAO, is the close partnership between academia and industry. This partnership has existed since the very beginning, when we began work on the concept design of the monitoring and control system. A possible area of weakness is in the development of a suitable talent pool of scientists and engineers who will enable optimal utilization of the telescope when it is ready. We have developed an ambitious human resource development program to make sure we have the right talent pool in place before full telescope operations commence around 2030.

**TMT and India-TMT**
The Thirty Meter Telescope (TMT) International Observatory (TIO) project is a next generation ground based optical and IR telescope facility mooted by institutes and science organizations in Canada, USA, Japan, China and India. There are two other large projects: the Giant Magellan telescope (GMT) and the European Extremely Large Telescope E-ELT). The project was expected to be completed in 2024 at Mauna Kea, Hawaii. Unfortunately, it didn't happen as the native The Hawaiians perceive the Mauna Kea as sacred. Since then, the project stakeholders and the local populations have been talking and discussing ways to accommodate both the science and culture. TMT once completed will provide breakthrough scientific discoveries across fields in astronomy: exoplanets, dark matter and dark energy, formation and evolution of planetary systems, and stellar and galactic systems. One of the key ingredients in the TMT project is the guaranteed in-kind contributions to the system by individual partner countries. This will foster technology cooperation and sharing the know-how among industry partners in the partner countries. India TMT engages many industries in meeting its in-kind obligations. Some systems are extra-ordinary complex which requires collective efforts across the partners to understand and provide solutions. India TMT has set-up a state-of-the-art optics fabrication facility in Bangalore with key inputs from other partner countries. The facility will start producing segments in late 2023 for the TMT project. The partnership in the large projects provide greater understanding of technical capabilities and social aspects in the partner countries.

India, as a member of the G20 and a participant in cutting-edge astronomy initiatives, is poised to provide significant scientific opportunities, particularly for nations in the global South. Collaborative projects like the Ground-based Next-Generation Optical and Infrared Telescopes (GMT, TMT, and E-ELT), the Square Kilometre Array (SKA), and the Laser Interferometer Gravitational-Wave Observatory (LIGO) exemplify the potential for transformative discoveries and technological innovation. The LIGO-India project, for instance, not only enhances our ability to pinpoint astrophysical events but also fosters multidisciplinary collaborations, facilitating knowledge exchange and building foundations for sustainable development. Similarly, India's involvement in SKA-India demonstrates its commitment to international partnerships, promoting unity, and sharing expertise across various domains. Close collaboration between academia and industry in India sets a precedent for effective cooperation and technology sharing. Additionally, the India-TMT initiative contributes to breakthroughs in astronomy while fostering technology cooperation among partner countries. India's active role in these large-scale projects showcases its commitment to offering scientific opportunities to G20 nations and the broader global South, advancing scientific progress, and promoting international cooperation.

**Recommendations**
1. Foster international collaboration and partnerships for global synergy.
2. Foster interdisciplinary collaboration to leverage expertise and innovation.
3. Foster industry-academia-government collaborations to create sustainable growth.
4. Secure sustained funding to ensure project continuity and impact.
5. Promote open data sharing and accessibility for widespread scientific collaboration.
6. Invest in advanced technologies to enhance data processing and analysis capabilities.
7. Facilitate technology transfer and collaborative R&D efforts through policy and incentives.
8. Prioritize sustainability and eco-conscious practices throughout project lifecycles.
9. Enhance education and workforce development for a skilled and diverse talent pool.
10. Promote diversity and inclusivity for a more comprehensive and equitable scientific community.
11. Define strategies to identify responsibilities and contributions within large scale projects with hundreds or thousands of collaborators.
12. Ease the travel burdens on the global South by facilitating visas for academic visits and conferences.

**Conclusions**
Large projects in astronomy are pivotal for advancing scientific knowledge and promoting sustainable development. By fostering international collaboration and partnerships, securing long-term funding, promoting open data sharing, and investing in technological advancements, these projects can maximize their scientific output, encourage innovation, and address global challenges. Prioritizing sustainability, enhancing education and workforce development, promoting diversity and inclusivity, and fostering interdisciplinary collaboration are key factors in ensuring the success and societal impact of these projects. By implementing these recommendations, we can harness the power of astroinformatics and large-scale astronomical endeavours to push the boundaries of our understanding of the universe while contributing to a more sustainable and inclusive future.

**S20 Co-Chair**: Ashutosh Sharma, Indian National Science Academy
**INSA S20 Coordination Chair:** Narinder Mehra, Indian National Science Academy

**Contributors**
Yogesh Wadadekar, National Center for Radio Astrophysics – TIFR, India
Shriharsh Tendulkar, Tata Institute of Fundamental Research, India
Sanjit Mitra, Inter-University Centre for Astronomy and Astrophysics, India
Eswar Reddy, Indian Institute of Astrophysics, India
A N Ramaprakash, Inter-University Centre for Astronomy and Astrophysics, India
Timo Prusti, European Space Agency (ESA), Netherlands
Juna Kollmeier, Canadian Institute for Theoretical Astrophysics, Canada
Pranav Sharma, Indian National Science Academy, India
Ashish Mahabal, California Institute of Technology, USA